\documentclass[]{jfm}

\usepackage{natbib}

\usepackage{amssymb}

\usepackage{url}

\usepackage{amsmath}
\usepackage{siunitx}

\usepackage{caption}
\usepackage{subcaption}

\usepackage{booktabs}
\usepackage[table]{xcolor}
\usepackage{booktabs}

\usepackage{listings}

\usepackage[capitalise]{cleveref}

\usepackage{placeins} 

\usepackage{pgf}
\usepackage{tikz}
\usetikzlibrary{arrows}

\usepackage{amsmath,amsfonts}
\usepackage{bm}

\newcommand{\im}{\mathrm{i}}
\newcommand{\cA}{\mathcal{A}}
\newcommand{\cB}{\mathcal{B}}
\newcommand{\cT}{\mathcal{T}}
\newcommand{\cW}{\mathcal{W}}
\newcommand{\cP}{\mathsf{P}}
\newcommand{\cM}{\mathsf{M}}
\newcommand{\cE}{\mathsf{E}}

\newcommand{\ttau}{\tilde\tau}
\newcommand{\bx}{\boldsymbol{x}}

\newcommand{\dd}[1]{\mathrm{d}#1}

\newcommand{\pa}{\partial}
\providecommand\der{{\rm d}}

\newcommand{\bbR}{\mathbb{R}}

\def\f#1#2{\frac{#1}{#2}}
\def\pp#1#2{\frac{\pa  #1}{\pa  #2}}
\def\dd#1#2{\frac{\der #1}{\der #2}}

\definecolor{lightred}{RGB}{255,122,105}
\definecolor{darkgreen}{RGB}{0,110,80}
\definecolor{darkpurple}{RGB}{140,0,140}

\tikzset{
    wow thick/.style={line width=2.1pt}
}

\tikzstyle{cst}=[circle,very thick,minimum size=6mm,draw=black]
\tikzstyle{end}=[circle,wow thick,minimum size=12mm,draw=black]
\tikzstyle{leaf}=[draw=darkgreen]
\tikzstyle{unav}=[draw=lightred]
\tikzstyle{merged}=[fill=blue!15]
\tikzstyle{fred}=[fill=red!30]
\tikzstyle{fblue}=[fill=blue!30]
\tikzstyle{fpurple}=[fill=darkpurple!30]
\tikzstyle{sbpic}=[->,>=stealth',auto,node distance=1.5cm]

\begin{document}

\title{The onset of filamentation on vorticity interfaces in two-dimensional Euler flows}

\author{David G. Dritschel\aff{1}\corresp{\email{david.dritschel@st-andrews.ac.uk}},
    Adrian Constantin\aff{2},
    Pierre M. Germain\aff{3}}

\affiliation{\aff{1}Mathematical Institute, University of St Andrews, St Andrews KY16 9SS, UK
\aff{2}Faculty of Mathematics, University of Vienna, 
Oskar-Morgenstern-Platz 1, 1090 Vienna, Austria
\aff{3}Department of Mathematics, Imperial College London, South Kensington Campus, 
London SW7 2AZ}

\maketitle

\begin{abstract}
Two-dimensional Euler flows, in the plane or on simple surfaces, possess a material invariant, namely the scalar vorticity normal to the surface. Consequently, flows with piecewise-uniform vorticity remain that way, and moreover evolve in a way which is entirely determined by the instantaneous shapes of the contours (interfaces) separating different regions of vorticity --- this is known as `Contour Dynamics'. Unsteady vorticity contours or interfaces often grow in complexity (lengthen and fold), either as a result of vortex interactions (like merger) or `filamentation'. In the latter, wave disturbances riding on a background, equilibrium contour shape appear to inevitably steepen and break, forming filaments, repeatedly --- and perhaps endlessly. Here, we revisit the onset of filamentation. Building upon previous work and using a weakly-nonlinear expansion to third order in wave amplitude, we derive a universal, parameter-free amplitude equation which applies (with a minor change) both to a straight interface and a circular patch in the plane, as well as circular vortex patches on the surface of a sphere. We show that this equation possesses a local, self-similar form describing the finite-time blow up of the wave slope (in a rescaled long time proportional to the inverse square of the initial wave amplitude). We present numerical evidence for this self-similar blow-up solution, and for the conjecture that almost all initial conditions lead to finite-time blow up. In the full Contour Dynamics equations, this corresponds to the onset of filamentation.
\end{abstract}

\begin{keywords}
{Inviscid two-dimensional flows, Contour Dynamics, filamentation}
\end{keywords}

{\bf PACS Codes} 47.35.-i, 47.32.C-

{\bf MSC Codes} {00A69, 76B47, 76M23, 37K05}


\section{Introduction}
\label{sec:intro}

Ideal two-dimensional flows, governed by Euler's equations, constitute an infinite-order Hamiltonian system \citep{Morrison1998}.  Not only are there energy and momentum invariants (symmetry permitting), but each fluid particle conserves its scalar vorticity \citep[the normal component of vorticity for general orientable surfaces, see][]{Saffman1995,Dritschel2015}. Because such flows are incompressible, they evolve purely by vorticity re-arrangement, determined by a velocity field that depends linearly, but non-locally, on the vorticity distribution. Few exact solutions are known, and almost all of them are either steady or in relative equilibrium (see the discussion in \cite{AbrashkinYakubovich,AlemanConstantin,ConstantinMartin,Crowdy,KWCC,MajdaBertozzi,Stuart}).

For piecewise-uniform vorticity distributions, the flow evolution depends only on the instantaneous shapes of the contours separating regions of uniform vorticity and on the vorticity jumps across them \citep{Zabusky1979,Dritschel1989}.  The resulting system of equations is called `Contour Dynamics'.  These equations also constitute an infinite-order Hamiltonian system, manifest in practice by contours which deform, elongate and generally grow in complexity, apparently indefinitely \citep{Dritschel1988f}. Nonetheless, mathematically, contours of a certain regularity class (at least possessing a continuous tangent vector) preserve that regularity class for all time \citep{Chemin1993,Bertozzi1993}.

The seemingly inevitable growth in complexity of these contours, or `vorticity interfaces', was first investigated by \cite{Dritschel1988f}, who studied the behaviour of small disturbances to circular contours (or `vortex patches') and to straight contours, both of which are otherwise in equilibrium. Numerical simulations performed using Contour Dynamics, including a regularisation called `surgery' \citep{Dritschel1988s}, demonstrated that a range of initial disturbances having small wave slope gradually steepened and folded over, a process called `filamentation'. Moreover, after the first filament forms, further filaments are generated at a frequency equal to half of the vorticity jump across the interface. For a fascinating historical account going back to Lord Kelvin \citep{Thomson1880}, see \cite{Craik2012}.

The present study focusses on the onset of filamentation, namely the wave-steepening stage and the approach to infinite wave slope. This was also studied mathematically in \cite{Dritschel1988f}, who developed a weakly-nonlinear theory describing the progressive steepening of an arbitrary disturbance. For this, the disturbance was expressed as a slowly-varying amplitude, $\cA$, multiplied by the relatively fast linear oscillation, $\exp(\im\omega t/2)$, with frequency equal to half of the vorticity jump ($\omega/2$) across the interface. Then. by expanding the equations of Contour Dynamics to third order in $\cA$, and by requiring no secular growth, a cubically-nonlinear evolution equation for $\cA$ was derived. This was shown to accurately describe the onset of filamentation by direct comparison with full Contour Dynamics numerical simulations.

In the present paper, we revisit this equation and demonstrate that, by introducing an appropriately re-scaled slow time, the amplitude equation for $\cA$ is parameter-free.  Hence, it applies equally well to disturbances propagating on circular contours in the plane and on the surface of a sphere.  It also applies, with a minor modification (one less term), to disturbances propagating on straight contours in the plane.  A companion paper \citep{Constantin2024} focuses on the mathematical structure of this equation, its conservation laws and symmetry properties, as well as exact solutions in special cases. Here, we study this equation numerically and, in particular, provide evidence for a finite-time singularity in the wave slope.  This singularity appears to have a self-similar form in the diffusive similarity variable $(x-x_c)/(t_c-t)^{1/2}$, where $x_c$ is the point of maximum wave slope and $t_c$ is the singularity time.

The paper is organised as follows. In \cref{sec:basics} we review the weakly-nonlinear theory presented in \cite{Dritschel1988f}, then show that the amplitude equation can be written in a parameter-free way in an appropriately re-scaled time variable. In \cref{sec:results}, we examine 
numerical simulations starting from a superposition of two low-wavenumber sinusoidal waves. Generically, for this and many other initial conditions, we find progressive wave steepening. Our results suggest a finite-time singularity, which is here analysed by fitting the wave shape to the form proposed in \cref{sec:results-sphere}. Our conclusions are offered in \cref{sec:discussion}.

\section{Weakly-nonlinear theory for vorticity interfaces}
\label{sec:basics}

We briefly recap the analysis presented in \cite{Dritschel1988f}, hereafter referred to as D88c.  In appendix A therein, a cubic-order amplitude equation was derived for small wave-slope disturbances to a circular vortex patch on the surface of a sphere. This in fact also applies to a circular patch in the two-dimensional plane $\bbR^2$ in a certain limit, as well as to a straight interface in $\bbR^2$, as explained below.

\subsection{The sphere}
\label{sec:sphere}

In Cartesian coordinates, the equations of Contour Dynamics on the sphere closely resemble those on the plane.  For a single contour $\mathcal{C}$ across which the vorticity jumps by $\omega$ (the difference between the vorticity to the left of the contour and that to the right), each point $\bx\in\mathcal{C}$ evolves according to
\begin{equation}
\label{CD}
    \dd{\bx}{t}=-\f{\omega}{4\pi}\oint_{\mathcal{C}}\log|\bx'-\bx|^2\,\der\bx'
\end{equation}
where $\bx'\in\mathcal{C}$ also \citep{Dritschel1988o}.  In the case of the sphere, the points
$\bx$ and $\bx'$ are three-dimensional but constrained to have magnitude $r=1$, without loss of generality. In the case of the plane, $\bx$ and $\bx'$ are two-dimensional. 

We now consider a circular vortex patch lying at constant latitude $\phi=\phi_0$, and separating vorticity $\omega_N$ to the north from $\omega_S$ to the south. Across the interface, the vorticity jumps by $\omega\equiv\omega_N-\omega_S$, which here we take to be unity, again without loss of generality. It is convenient to consider displacements in the axial coordinate $z=\sin\phi$ of the vortex patch boundary, in particular displacements of the form
\begin{equation}
\label{zform}
    z(\theta,t)=z_0-r_0^2\rho(\theta,t)
\end{equation}
where $\theta$ is the longitude coordinate, $t$ is time, $z_0=\sin\phi_0$ and $r_0=\cos\phi_0$.  Here $\rho(\theta,t)$ is the displacement function, considered small compared to unity (this form facilitates taking the planar limit, i.e.\ $\phi_0\to\pi/2$).  We follow (A6) in D88c and take
\begin{equation}
\label{rform}
    \rho(\theta,t)=a\eta(\theta,t)
\end{equation}
where $a\ll 1$ and $\eta={\mathcal{O}}(1)$.  We assume here that the mean value of $\rho$, and therefore $\eta$, vanishes.  (The mean value plays no role in the disturbance evolution, and corresponds to an ${\mathcal{O}}(a)$ shift of the mean latitude $\phi_0$.)

\begin{figure}
    \centering
    \includegraphics[width=0.75\textwidth]{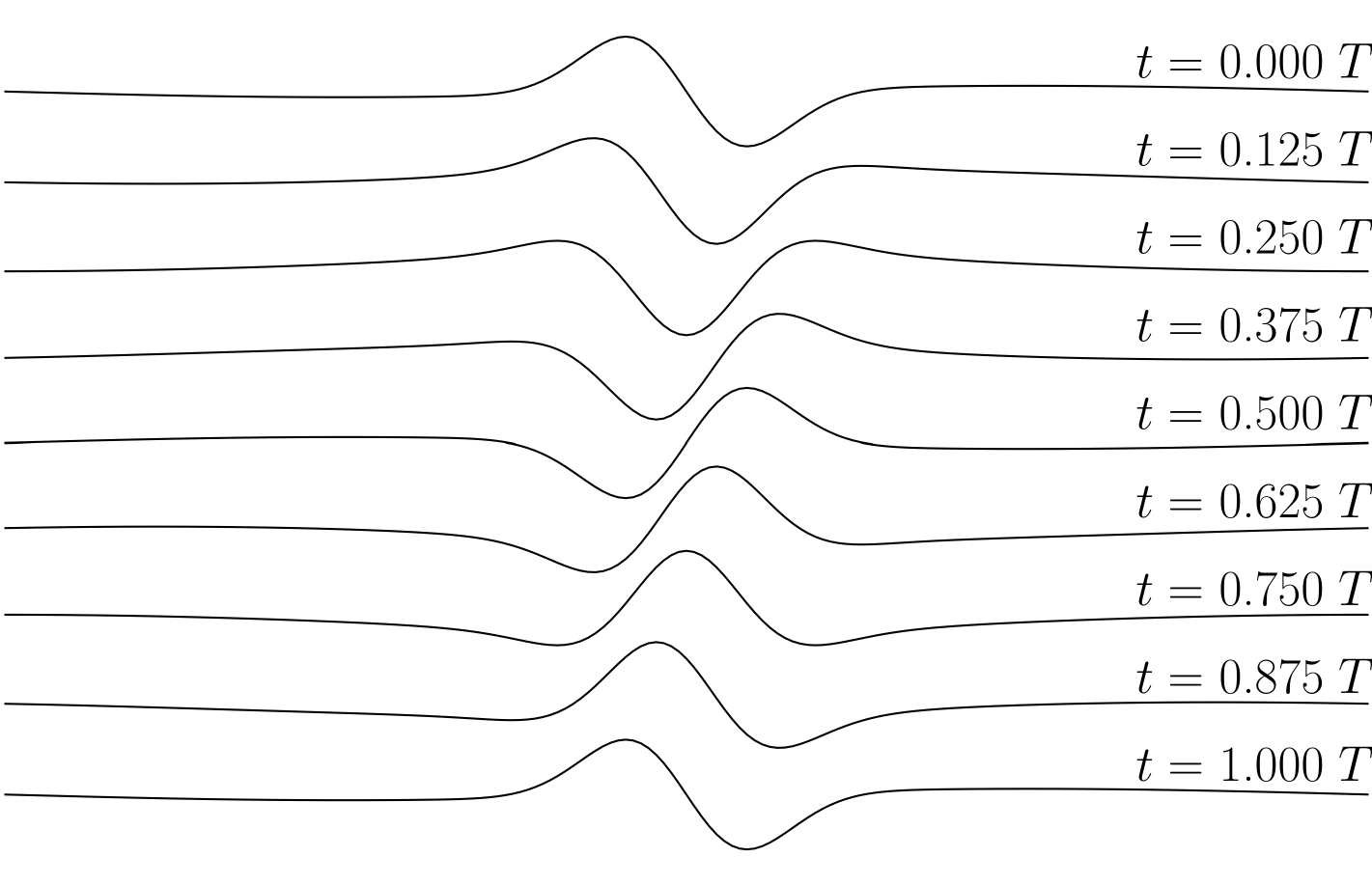}
    \caption{Illustration of the linear evolution of disturbances to a vorticity interface. Here, $T=4\pi/\omega$ is the linear wave period. Note that the initial disturbance reverses after half a period, then recovers its initial form after one full period. After a quarter of a period, the solution turns from anti-symmetric to symmetric about its centre, and this also reverses after a further half period.}
    \label{fig:hilbert}
\end{figure}

Expanding \eqref{CD} to ${\mathcal{O}}(a^3)$, the equation satisfied by $\eta$ takes the
form
\begin{align}
    \frac{1}{\omega}&\pp{\eta(\theta,t)}{t}-\frac{1}{4\pi}\int_0^{2\pi}\frac{\eta(\alpha,t)\sin(\alpha-\theta)}{1-\cos(\alpha-\theta)}\der\alpha=\pp{G}{\theta} \nonumber \\
    &G(\theta,t)=\frac{a}{4}z_0\eta^2-\frac{a^2}{6}(z_0^2+1)\eta^3
    -\frac{a^2}{24\pi}\int_0^{2\pi}\frac{[\eta(\alpha,t)-\eta(\theta,t)]^3}{1-\cos(\alpha-\theta)}\der\alpha\,,
    \label{cubiceq}
\end{align}
in a frame of reference rotating with the mean angular velocity $\tfrac12\omega$ of the patch boundary (this is (A7) in D88c, for zero mean $\eta$). Without the nonlinear term ($G=0$), the equation is linear in $\eta$, and all solutions oscillate at the common frequency $\tfrac12\omega$, \textit{independent} of the spatial structure of $\eta$.  Specifically, in linear theory, the general solution to \eqref{cubiceq} for $G=0$ is
\begin{equation}
    \label{linear}
    \eta(\theta,t)=\cA(\theta)e^{\shalf\im\omega t} + \mathrm{c.c.} \quad\textrm{with}\quad
    \cA(\theta)=\sum_{m=1}^{\infty}a_m e^{\im m\theta}
\end{equation}
where the coefficients $a_m$ are arbitrary complex constants, and `c.c.' denotes complex conjugation.  Denoting the real and imaginary parts of $\cA$ by $\cA_r$ and $\cA_i$ respectively, we can write this solution in the form 
\[
\eta(\theta,t)=2\left(\cA_r(\theta)\cos(\thalf\omega t)-
                      \cA_i(\theta)\sin(\thalf\omega t)\right)\,.
\]
This is illustrated in figure \ref{fig:hilbert} for an initially anti-symmetric disturbance.  Note that by a quarter of the linear wave period, the solution has become symmetric (this is $-2\cA_i$), then changes back to the initial disturbance except reversed after another quarter period, and so on.  The upshot is that the linear evolution is dynamic, evolving on a time scale inversely proportional to the vorticity jump $\omega$.

For the nonlinear equation \eqref{cubiceq} with $G\neq0$, we follow the multiple-time-scales ansatz in (A8) of D88c and look for a solution of the form
\begin{equation}
\label{eform}
    \eta(\theta,t)=\cA(\theta,t)e^{\shalf\im\omega t} + \mathrm{c.c.}
\end{equation}
where $\cA(\theta,t)=\cA_0(\theta,\tau)+a\cA_1(\theta,t,\tau)+...$, and where $\tau=\omega a^2 t$ is the slow time.  Omitting the details, the equation for $\cA_0$ (hereafter written simply as $\cA$) is (A11) of D88c, rewritten here as
\begin{equation}
\label{aeq1}
    \pp{\cA}{\tau}=\f12 \pp{}{\theta}
    \left(z_0^2 \cT_1 + \cT_2 - (z_0^2+1)|\cA|^2\cA\right)
\end{equation}
where
\begin{equation}
\label{t1}
    \cT_1=\im\left(\cA\pp{}{\theta}(\cW-\bar{\cW})-|\cA|^2\pp{\cA}{\theta}\right)
\end{equation}
and
\begin{equation}
\label{t2}
    \cT_2=-\f{1}{4\pi}\int_0^{2\pi}
    \frac{|\cA(\alpha,\tau)-\cA(\theta,\tau)|^2[\cA(\alpha,\tau)-\cA(\theta,\tau)]}
    {1-\cos(\alpha-\theta)}\,\der{\alpha}\,.
\end{equation}
Above, $\cW$ is the part of $|\cA|^2$ expressible in positive wavenumbers, i.e.
\[
\cW=\sum_{m=1}^{\infty} w_m e^{\im m\theta} \,,
\]
and an overbar denotes complex conjugation (note that $|\cA|^2=\cW+\bar{\cW}+\cP$, where $\cP$ is a constant of the motion, see below). By explicit calculation starting with the general Fourier series 
\[
\cA(\theta,\tau)=\sum_{m=1}^{\infty}a_m(\tau) e^{\im m\theta} \,,
\] 
remarkably, it can be shown that $\cT_2=\cT_1$ \citep{Constantin2024}.  In this case, by introducing the re-scaled slow time $\ttau=\thalf(z_0^2+1)\tau$, we have more simply
\begin{align}
    \label{aeq2}
    \pp{\cA}{\ttau} &=\pp{\cB}{\theta} \\
    \cB(\theta,\ttau)&=-\f{1}{4\pi}\int_0^{2\pi}
    \frac{|\cA(\alpha,\ttau)-\cA(\theta,\ttau)|^2[\cA(\alpha,\ttau)-\cA(\theta,\ttau)]}
    {1-\cos(\alpha-\theta)} \,\der{\alpha}
    -|\cA(\theta,\ttau)|^2\cA(\theta,\ttau) \nonumber \,,
\end{align}
where we have chosen $\cB=\cT_2-|\cA|^2\cA$ above.

Equation \eqref{aeq2} models the onset of filamentation on circular vortex patches at \textit{any} `height' $z_0$, \textit{including} the planar limit $z_0\to1$. In particular, it does not explicitly depend on the values of the vorticity $\omega_N$ and $\omega_S$ to the north and south of the interface.  Of course, the vorticity jump requires $\omega_N-\omega_S=\omega$, and the vanishing of the mean vorticity over the sphere (by Stokes' theorem) requires $(1-z_0)\omega_N+(1+z_0)\omega_S=0$. Combining these relations gives
\begin{equation}
\label{omegavalues}
    \omega_N=\tfrac12(1+z_0)\omega \qquad \& \qquad
    \omega_S=-\tfrac12(1-z_0)\omega \,,
\end{equation}
showing that, in general, the average vorticity near the interface,
\begin{equation}
\label{omegaavg}
    \tfrac12 (\omega_N+\omega_S)=z_0\omega \,,
\end{equation}
depends on $z_0$.  This average vorticity induces a mean shear across the patch boundary unless $z_0=0$, yet this shear plays no role in the filamentation equation \eqref{aeq2} above, apart from changing the definition of the dimensionless time $\ttau$.

There is also a spectral form of this equation.  When deducing $\cT_1=\cT_2$ above, we found that the Fourier coefficients $a_m(\ttau)$ of $\cA(\tilde\theta,\ttau)$ evolve according to 
\begin{equation}
\label{eqaspec}
    \dd{a_m}{\ttau}=\tfrac12 \im m \sum_n\sum_p (n+p-|n-m|-|p-m|-2)a_n a_p
    \bar{a}_{n+p-m} \,.
\end{equation}
(This corrects (A12) in D88c, where the term involving $z_0^2$ there is incorrect.) In \eqref{eqaspec}, the sums over $n$ and $p$ are restricted to integers satisfying $n+p>m$, for each $m$.  Notably, $n+p-|n-m|-|p-m|-2=2(m-1)$ whenever both $n\geq m$ and $p\geq m$.  In particular, this shows that $a_1$ is an invariant of the dynamics: $\der a_1/\der{\ttau}=0$. It can be shown that this invariant corresponds to conservation of the $x$ and $y$ components of the angular impulse vector on the sphere,
\begin{equation}
    \label{angimp}
    {\bm{J}}=\f{1}{2}\omega\oint_{\mathcal{C}} {\bm{x}}\times\der{\bm{x}}
\end{equation}
\citep[see Appendix B of][]{Polvani1993}.
Additionally, the `momentum' $\cP$, `mass' $\cM$ and (kinetic) energy $\cE$ are all invariant \citep{Constantin2024}; the momentum and mass are simple sums over spectral coefficients,
\begin{equation}
\label{invariants}
    \cP=\sum_{m>0} |a_m|^2 \quad\textrm{and}\quad \cM=\sum_{m>0} |a_m|^2/m \,,
\end{equation}
while the energy has a more complex form \citep[see][for details]{Constantin2024}.

\subsection{The line}
\label{sec:line}

In an appropriate limit, \eqref{aeq2} (or \eqref{eqaspec}) also applies to the onset of filamentation for weakly-nonlinear disturbances to a straight interface, say $y=0$, on $\bbR^2$. Guided by the analysis in Appendix B of D88c, we substitute $\theta=-\kappa x$, $\alpha=-\kappa x'$ and $\cA=\kappa \hat{\cA}$ into \eqref{aeq2}. Then, we take the limit $\kappa \to 0$, assuming $|\alpha-\theta| \ll 1$. After dividing by $\kappa$ and dropping the hat on $\hat{\cA}$, we find
\begin{align}
    \label{aeql}
    \pp{\cA}{\ttau}&=\pp{\cB}{x} \\
    \cB(x,\ttau)&=\f{1}{4\pi}\int_0^{2\pi}
    \frac{|\cA(x',\ttau)-\cA(x,\ttau)|^2
    [\cA(x',\ttau)-\cA(x,\ttau)]}
    {1-\cos(x'-x)}\,\der{x'} \nonumber \,,
\end{align}
assuming periodicity in $x$ on the interval $[0,2\pi]$. Without periodicity, one would need to replace $1-\cos(x'-x)$ by $\tfrac12(x'-x)^2$ and extend the integration over $x'$ to the whole real line (this is effectively (B18) in D88c). This equation has also been re-derived by \cite{BH}, but in the differential form involving $\cT_1$ in \eqref{t1} (with $\theta$ replaced by $-x$) --- see \cite{Constantin2024} for further details.

The only important difference from the spherical case is the absence of the `curvature' term $|\cA|^2\cA$. There is also an unimportant sign change due to changing from spherical to planar Cartesian coordinates. The mode amplitudes (or Fourier coefficients) $a_k(\ttau)$ in 
\[
\cA(x,\ttau)=\sum_{k=1}^{\infty} a_k(\ttau) e^{\im k x}
\]
evolve according to
\begin{equation}
\label{leqaspec}
    \dd{a_k}{\ttau}=-\tfrac12 \im k \sum_n\sum_p (n+p-|n-k|-|p-k|)a_n a_p
    \bar{a}_{n+p-k} \,.
\end{equation}
Compared with \eqref{eqaspec}, the `$-2$' in the bracket there is now missing --- this came from the curvature term $|\cA|^2\cA$ in the circular case.  Now, $a_1$ in general varies in time: none of the $a_k$ are invariant. However, momentum $\cP$ and mass $\cM$ as defined in \eqref{invariants} are still invariant, as is the energy $\cE$ \citep[see][]{Constantin2024}.

\subsection{Numerical approach}
\label{sec:method}


Three mathematically-equivalent forms of the equations were discretised (or truncated) to form a finite system of independent variables. Here we discuss the codes for the sphere, but the codes for the line are closely similar (they only lack the curvature term $|\cA|^2\cA$). The codes, at early times, give results which are indistinguishable from one another.

All codes use a standard fourth-order Runge-Kutta time integration scheme, with time-step adaptation. The time step is limited to 
\begin{equation}
\label{timestep}
  \Delta\ttau=\min\left(\alpha/r_{\max},\,\beta/M,\,\ttau_{\mathrm{save}}-\ttau\right) 
  \quad\textrm{with}\quad r_{\max}=|\pa^2{\cA_r}/\pa{\ttau}\pa{\theta}|_{\max} \,,
\end{equation}
where $\ttau_{\mathrm{save}}$ is the next data save time and $\cA_r$ is the real part of $\cA$. Here we have chosen $\alpha=0.2$ (or $\alpha=0.1$ in the case of the line whose evolution is faster), but results with half this value are almost indistinguishable. The time step $\alpha/r_{\max}$ attempts to resolve the increasingly high frequencies on the approach to any singularity. We have also chosen $\beta=2048\times10^{-3}$; the time step $\beta/M$ acts as a CFL constraint. These criteria for limiting the time step were deduced by trial and error for a wide range of initial conditions.  We found that it is insufficient to use simply $r_{\max}=|\pa{\cA_r}/\pa{\ttau}|_{\max}$ --- then spurious high wavenumber ($m$) noise develops and eventually visibly contaminates the solution.

We first developed a pseudo-spectral code for \eqref{aeq2} but using the differential form $\cB=\cT_1-|\cA|^2\cA$ in \eqref{t1}. (Note that it is sufficient to evolve only the real part of $\cA$, denoted $\cA_r$, since the imaginary part, $\cA_i$, can be deduced from the Fourier coefficients of $\cA_r$.) In this pseudo-spectral code, all nonlinear products are computed on a uniform grid of $N=2M$ points, while derivatives are calculated spectrally, making use of Fast Fourier Transforms (FFTs). Due to the cubic nonlinearity, de-aliasing was applied by removing the upper half of all wavenumber coefficients before computing products on the grid. We also tried the `8-pole' Butterworth filter $(1+(m/m_c)^{16})^{-1/2}$ with $m_c=M/2$ (half the maximum wavenumber).  However, neither filter was sufficient to stabilise the code over the integration times of interest. 

Second, we developed a code for \eqref{aeq2} directly.  Here, the nonlinear products are computed with a filtered amplitude $\tilde{\cA}$ in place of $\cA$ (using the same Butterworth filter above). Derivatives are computed spectrally (after FFTs). This code proves to be numerically stable only if the curvature term $|\cA|^2\cA$ is omitted (this applies to the line case). Otherwise spurious high-wavenumber noise develops which we have been unable to suppress by filtering. 

Third, we solved for the spectral coefficients directly, using \eqref{eqaspec}, with the sums over $n$ and $p$ truncated to ensure $n$, $p$ and $n+p-m$ all lie in the range $[1,M]$. This is much more computationally intensive than the pseudo-spectral code, but it is stable over the times of interest. However, numerical stability requires the introduction of a damping term on the right-hand side of the evolution equation \eqref{aeq2}. In spectral space, this has the form $-\nu\,(m/M)^p a_m$ in \eqref{eqaspec}, where 
\begin{equation}
    \nu(\ttau)=\frac{\varepsilon\,r_{\max}(\ttau)}{\alpha} \,. 
\end{equation}
By experimentation, we have found that choosing $p=6$ and $\varepsilon=10^{-2}$ ensures that the upper end of the variance spectrum $|a_m|^2$ remains small compared to lower and mid-range over the entire simulation period. In effect, the highest wavenumber is damped by the factor $e^{-\varepsilon}$ every time step when the time step is limited by $r_{\max}$.  Notably, this damping did not stabilise either of the other two codes developed.

\section{Results}
\label{sec:results}

\subsection{The sphere}
\label{sec:results-sphere}

We focus on one example in this case which nonetheless typifies the behaviour seen in many others.  The amplitude $\cA$ is initialised with only two non-zero spectral coefficients, specifically $a_2=\sqrt{3}/2$ and $a_3=0.25e^{\im\pi/3}$. Then, the `momentum' $\cP=|a_2|^2+|a_3|^2$ equals $1$, and $\cP=\sum_m |a_m|^2$ remains equal to $1$ for all time $\ttau$.  We have performed simulations with mode truncations $M=256$, $512$, $1024$ and $2048$.  All give similar results, but the highest resolution allows a better resolution of the approach to a singularity in wave slope.

\begin{figure}
    \centering
    \includegraphics[width=\textwidth]{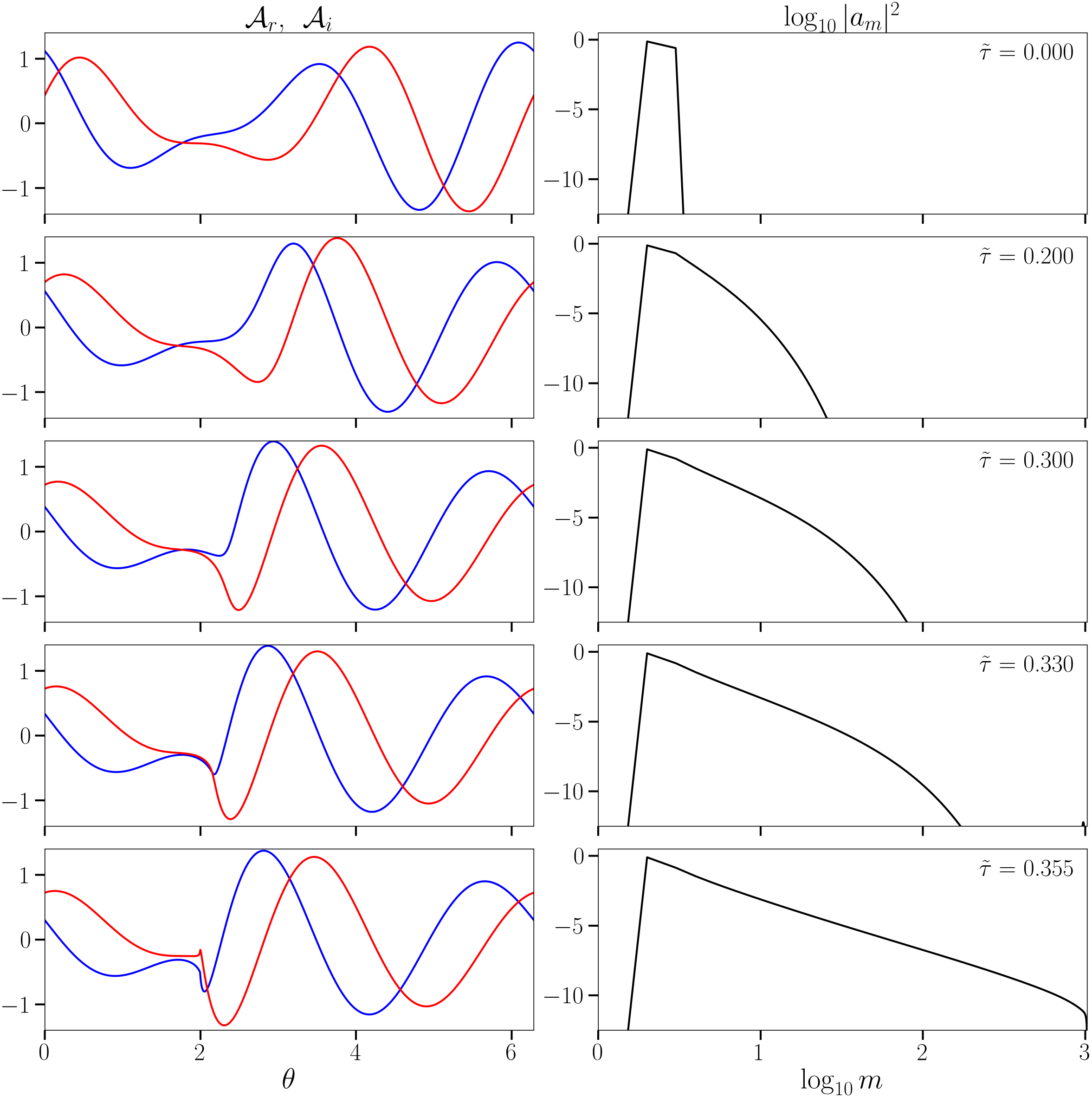}
    \caption{Time evolution of the wave amplitude $\cA$ (left, with real and imaginary parts in blue and red respectively) for a circular vorticity interface on a sphere, together with the corresponding power spectrum $|a_m|^2$ (right) for 5 selected times $\ttau$ (increasing downwards).  Initially, just $a_2$ and $a_3$ are non-zero. [colour online]}
    \label{fig:circle_evo}
\end{figure}

\begin{figure}
    \centering
    \includegraphics[width=0.625\textwidth]{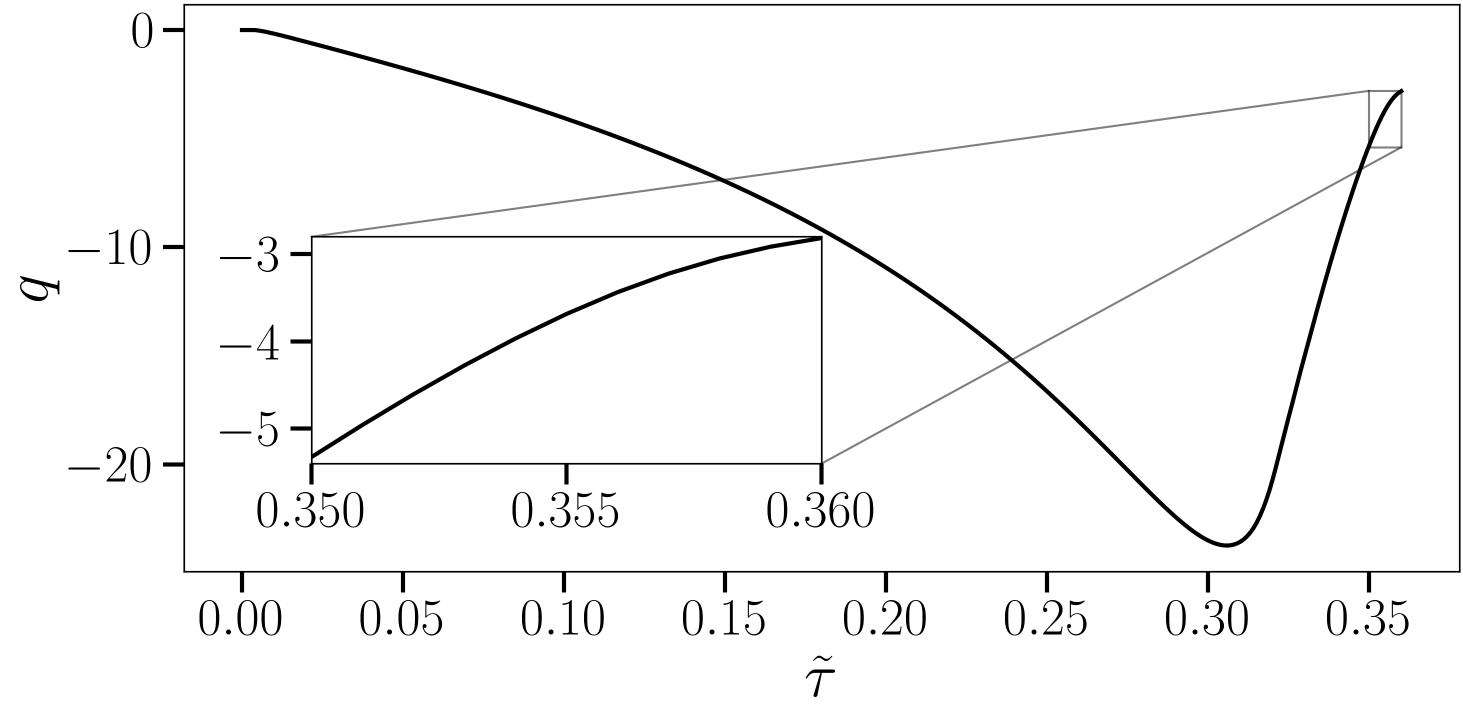}
    \caption{Time evolution of the spectral slope $q$, obtained by a least-squares fit of $\log|a_m|^2$ to $q\log{m}+c$, between wavenumbers $m=10$ and $M/2=1024$.  The inset shows the slope over the last $0.1$ units of time computed (note, the results are not considered reliable beyond $\ttau=0.355$ due to insufficient resolution).}
    \label{fig:circle_slope}
\end{figure}

The evolution of $\cA$ and of its spectrum are shown at a few selected times in figure \ref{fig:circle_evo}. The left column shows the real and imaginary parts of the amplitude $\cA(\theta,\ttau)$, while the right column shows the spectrum $|a_m(\ttau)|^2$.  In general, the waves comprising the interface propagate to the left, but they also steepen. In the spectrum, this is associated with the development of a powerlaw form $|a_m|^2 \propto m^{q}$, with $q\approx -3$, over a large range of wavenumbers $m$ --- see figure \ref{fig:circle_slope}. At the final reliable time shown ($\ttau=0.355$), the amplitude $\cA$ exhibits a near discontinuous variation, which is nonetheless marginally resolved (as seen e.g.\ in a zoom at this time, discussed below).  

\begin{figure}
    \centering
    \includegraphics[width=\textwidth]{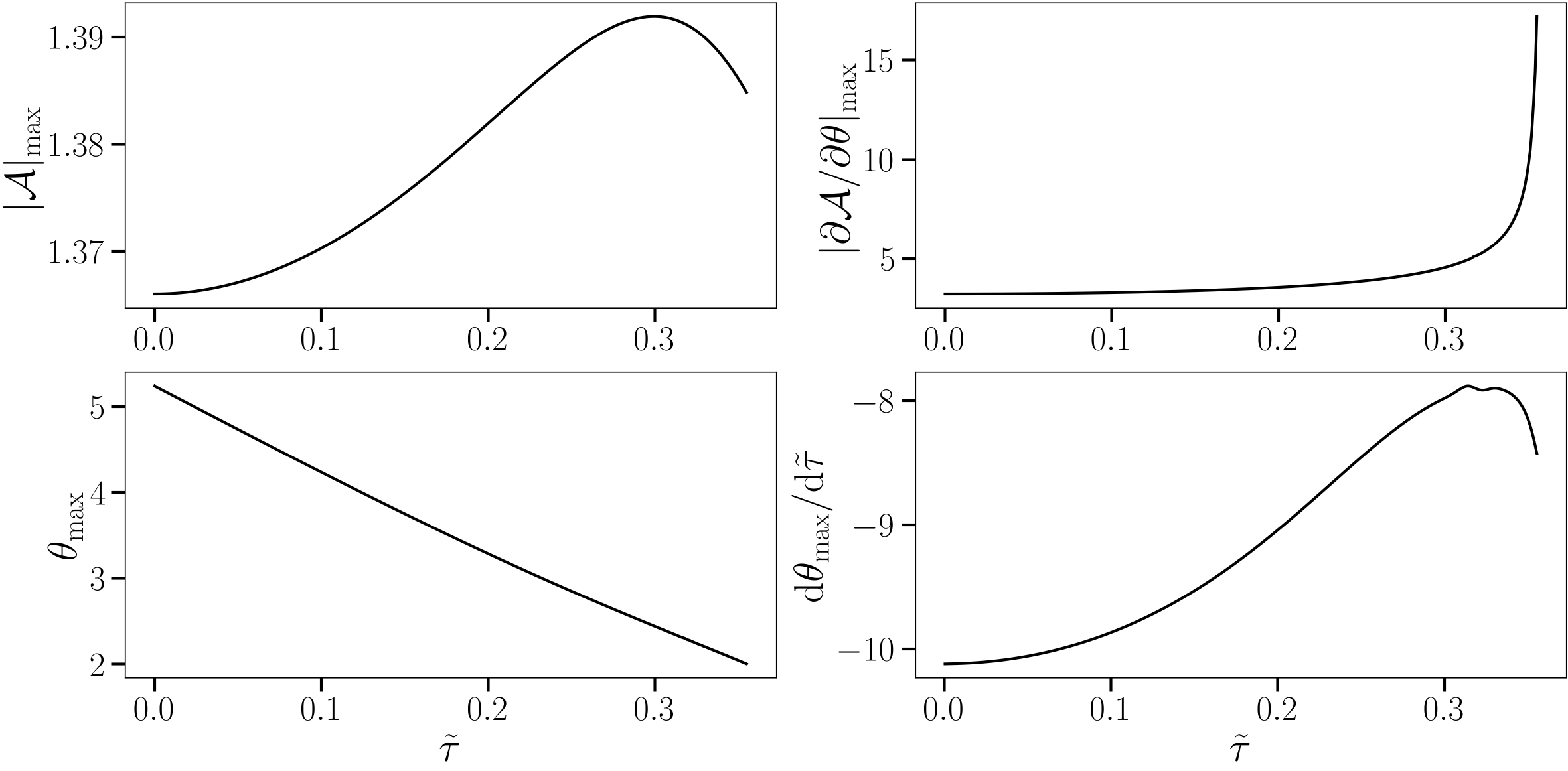}
    \caption{Time evolution of various diagnostics, as labelled, from $\ttau=0$ to $0.355$ (the last reliable time). In the figure for $\der\theta_{\max}/\der\ttau$, 1-2-1 averages were repeated 64 times to remove most of the noise occurring around the maximum near $\ttau=0.32$ (endpoint values were replaced by linear extrapolation of the averaged interior data points).  This noise arises from the imprecision in locating $\theta_{\rm max}$.}
    \label{fig:circle_diag}
\end{figure}

We next provide evidence that the solution is approaching a finite-time singularity in the 
wave slope $|\partial\cA/\partial\theta|_{\max}$. This wave slope is determined by finding the $\theta$ grid point having the largest value of $|\partial\cA/\partial\theta|$, then fitting a parabola through this value and the values on either side of this grid point. The maximum in this parabola occurs at $\theta=\theta_{\rm max}$, which varies with $\ttau$. The same procedure is also used to determine the largest value of $|\cA|$, to better understand the nature of the solution near the conjectured singularity. These diagnostics are shown in figure ~\ref{fig:circle_diag}. One sees a dramatic rise in the maximum wave slope (upper right panel) near the conjectured singularity time, whereas other diagnostics are tending to finite values.  In particular, the maximum amplitude $|\cA|_{\rm max}$ hardly varies over the entire evolution. Also, the location of maximum wave slope moves at a moderate speed, between $-10$ and $-8$ units approximately.

\begin{figure}
    \centering
    \includegraphics[width=\textwidth]{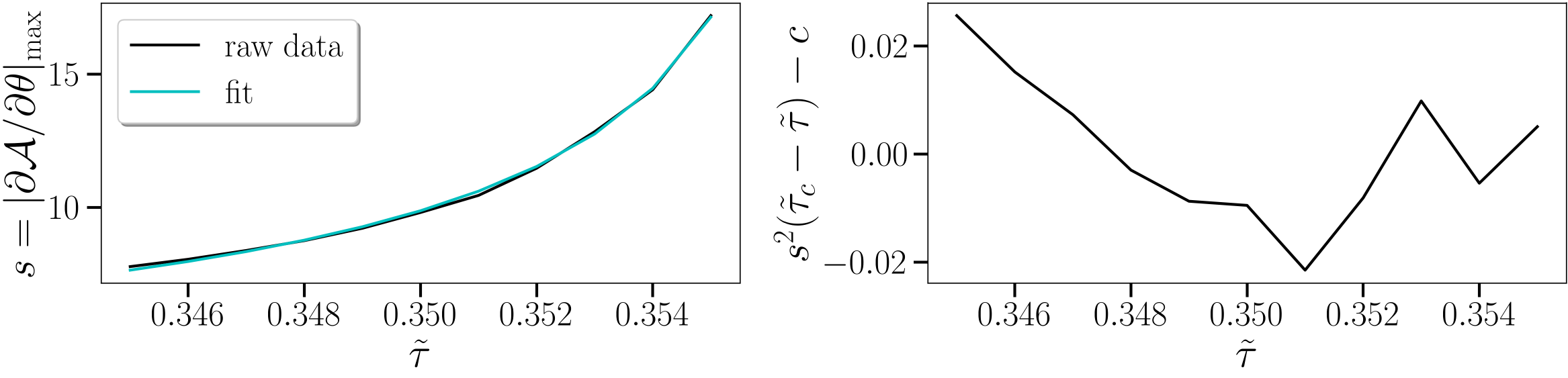}
    \caption{Time evolution of the maximum wave slope $s(\ttau)=|\partial\cA/\partial\theta|_{\max}$ together with a fit to $\sqrt{c/(\ttau_c-\ttau)}$ (left), and the function $f(\ttau)=s^2(\ttau_c-\ttau)-c$ (right) which would be zero for a perfect fit. [colour online]}
    \label{fig:circle_tfit}
\end{figure}

The maximum wave slope $s(\ttau)=|\partial\cA/\partial\theta|_{\max}$ appears to exhibit a square-root singularity, i.e.\ $s\sim\sqrt{c/(\ttau_c-\ttau)}$ for $\ttau_c\approx0.35748$, as shown in figure \ref{fig:circle_tfit}. The singularity time $\ttau_c$ and the constant $c$ were determined by a least-squares fit over the period $0.345\leq\ttau\leq0.355$. Specifically, $\ttau_c$ and $c$ were determined by minimising $\sum w(\ttau_j)f^2(\ttau_j)$ for discrete times $\ttau_j$, where $f=s^2(\ttau_c-\ttau)-c$ and $w=s$, a weight chosen to favour data closer to the singularity time. The r.m.s.\ error in the fit over this period is only $0.01189$ (note: $c\approx0.72935$). 

\begin{figure}
    \centering
    \includegraphics[width=0.625\textwidth]{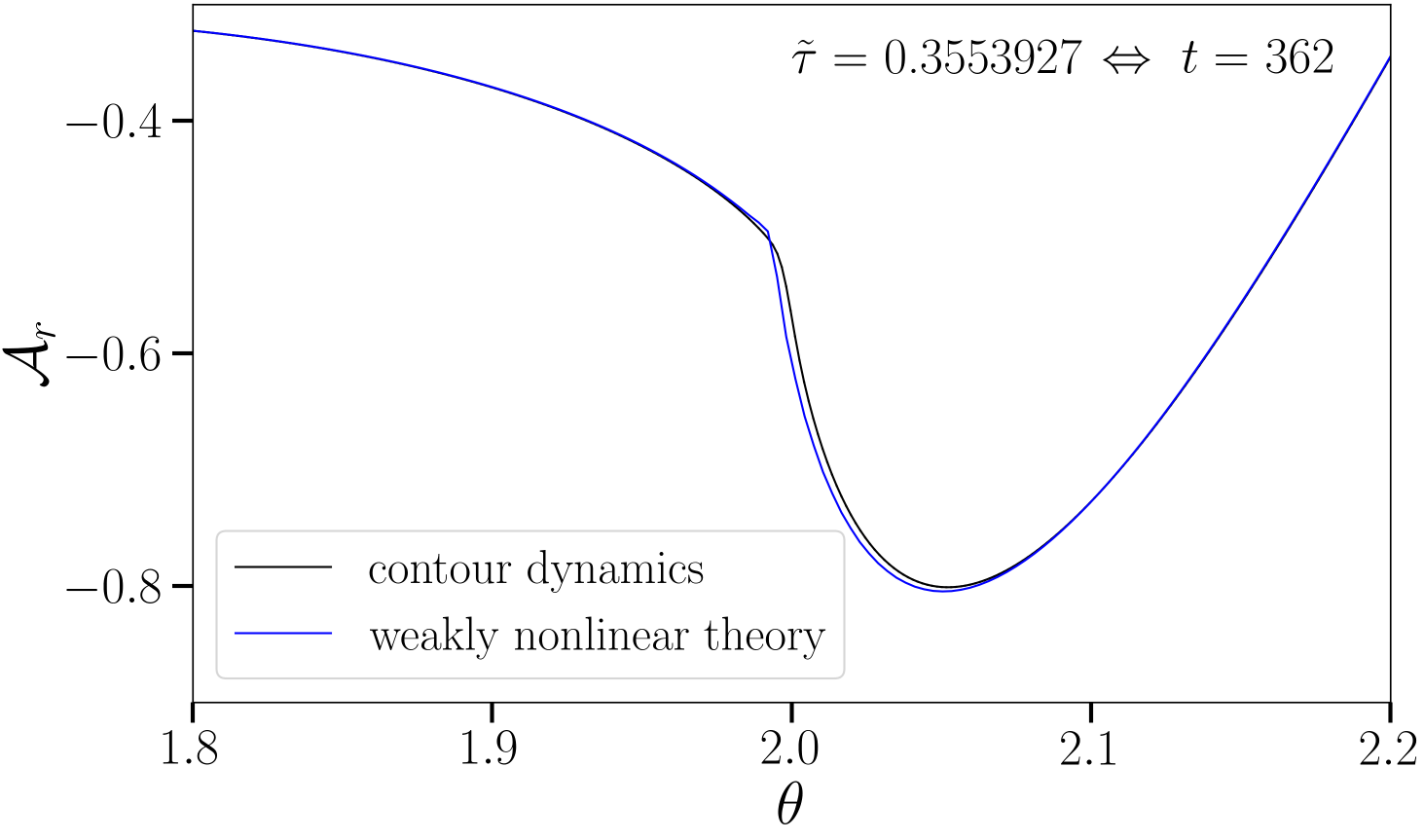}
    \caption{Zoom of a portion of the curve $\cA_r(\theta,\ttau)$ at $\ttau\approx0.3553927$ (blue) compared with a contour dynamics simulation (black) at the equivalent time, here $362$ linear wave periods (see text for details). [colour online]}
    \label{fig:compare}
\end{figure}

To appreciate how well the weakly nonlinear theory captures the onset of filamentation, it is compared in figure \ref{fig:compare}, at a time just beyond the last time shown in figure \ref{fig:circle_evo}, with the results of a full contour dynamics simulation solving \eqref{CD} numerically. The contour dynamics simulation was initialised with $z_0=0$ (i.e.\ the equator) and $\rho(\theta,0)=2a\cA_r(\theta,0)$ in \eqref{zform}, with amplitude $a=1/40$. The vorticity jump across the interface was taken to be $\omega=4\pi$ so that one unit of time corresponds to one full linear wave oscillation. The numerical parameters are now standard and may be found in \cite{Dritschel2010}, apart from the large-scale length $L$ used to control the point resolution (here we take $L=6.25\pi/M\approx0.009587$, giving 3936 initial contour nodes).  The agreement in figure \ref{fig:compare} is excellent, apart from a small discrepancy near the point of maximum wave slope. At earlier times, the solutions overlap within the plotted line width.

\begin{figure}
    \centering
    \includegraphics[width=\textwidth]{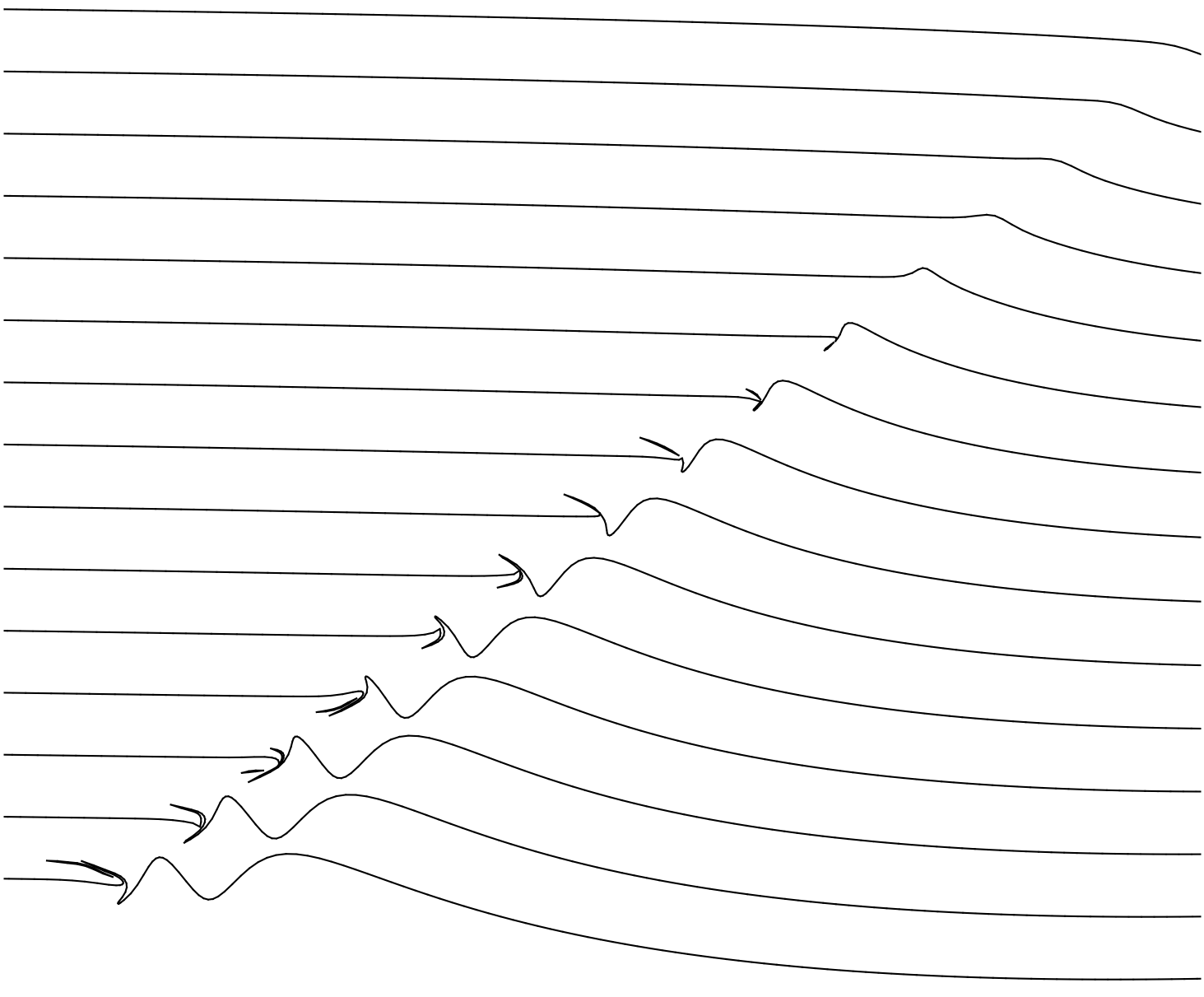}
    \caption{A portion of the interface in the full contour dynamics simulation at successive linear wave periods, from $t=362$ at the top to $t=376$ at the bottom. Here, $\rho(\theta,t)$ is plotted over the range $1.85\leq\theta\leq2$. Note, that between the times shown, the interface exhibits a relatively fast oscillation over the linear wave period, analogous to that illustrated in figure \ref{fig:hilbert}.}
    \label{fig:contours}
\end{figure}

Shortly after this time, the waves on the interface do break in the contour dynamics simulation, as shown in figure \ref{fig:contours}. Filamentation starts between $t=366$ and $367$ (note the interface shape changes greatly \textit{between} each period, due to the underlying linear oscillation). In terms of the long time scale, filamentation occurs between $\ttau=0.3593$ and $0.3603$,
a little later than the time $\ttau_c\approx0.35748$ indicated by the singularity fit in the weakly nonlinear theory.

The behaviour seen in this example appears to be common to any (non-periodic) initial condition: inevitably the maximum wave slope appears to blow up in finite time, and this is associated with the spectrum filling out, possibly tending to the form $|a_m|^2 \sim m^{-3}$ at the singularity time. Moreover, the blow up is local, occurring over a small range in $\theta$. Motivated by these observations, we seek an approximate solution to \eqref{aeq2} that exhibits a local, self-similar, finite-time singularity. To this end, we introduce the similarity variable
\begin{equation}
\label{ssvar}
    \xi=\frac{\theta-\theta_{\rm max}(\ttau)}{\sqrt{\delta}} \quad\textrm{with}\quad
    \delta=1-\ttau/\ttau_c
\end{equation}
and seek an approximate solution of the form
\begin{equation}
\label{ssform}
    \cA(\theta,\ttau)=\delta^{\im \mu} \psi(\xi) \,,
\end{equation}
for some constant $\mu$. This is the only self-similar form consistent with the equations of motion: the imaginary exponent of $\delta$ in the pre-factor guarantees that $|\cA|_{\max}$ does not vary significantly near the singularity time, as observed in figure \ref{fig:circle_diag}. Moreover, this form leads to $|\partial\cA/\partial\theta|_{\max}\propto1/\sqrt{\ttau_c-\ttau}$, again as observed.

An equation for $\psi(\xi)$ can be obtained as follows.  Take $\theta=\theta_{\max}+\delta^\shalf\xi$ and $\alpha=\theta_{\max}+\delta^\shalf\xi'$ in \eqref{aeq2}. Then, assuming $\delta\ll1$ while $\xi'-\xi=\mathcal{O}(1)$, one can show that $\cB=\delta^{\im\mu-\shalf}b(\xi)$ to leading order, where
\begin{equation}
\label{ssbeq}
    b(\xi)=-\f{1}{2\pi}\int_{-\infty}^{\infty}
    \frac{|\psi(\xi')-\psi(\xi)|^2
    [\psi(\xi')-\psi(\xi)]}
    {(\xi'-\xi)^2}\,\der{\xi'} 
\end{equation}
(assuming sufficiently fast far-field decay of the integrand). Notably, the curvature term $|\cA|^2\cA$ in \eqref{aeq2} is $\mathcal{O}(\delta^{\shalf})$ smaller, so is neglected here. The right-hand side of the evolution equation \eqref{aeq2}, namely $\partial\cB/\partial\theta$, is thus $\delta^{\im\mu-1}\der{b}/\der{\xi}$ to leading order.  The left-hand side evaluates to
\begin{equation}
    \im\mu\delta^{\im\mu-1}\psi+
    \delta^{\im\mu}\frac{\der\psi}{\der\xi}\frac{\partial\xi}{\partial\ttau} \,.
\end{equation}
However, since $\xi=(\theta-\theta_{\max})\delta^{-\shalf}$, we have
\begin{equation}
\label{partial}
    \frac{\partial\xi}{\partial\ttau}=-\delta^{-\shalf}\frac{\der\theta_{\max}}{\der\ttau}+\delta^{-1}\frac{\xi}{2\ttau_c} \,.
\end{equation}
For $\delta\ll1$, we can neglect the $\mathcal{O}(\delta^{-\shalf})$ term, implying
that the left-hand side of \eqref{aeq2} is
\begin{equation}
    \delta^{\im\mu-1}\left(\im\mu\psi+\frac{\xi}{2\ttau_c}\frac{\der\psi}{\der\xi}\right)
\end{equation}
to leading order.  This has the same $\delta^{\im\mu-1}$ pre-factor as the right-hand side,
so upon cancelling this pre-factor we obtain an equation entirely in terms of $\xi$:
\begin{equation}
\label{sseq}
    \im\mu\psi+\frac{\xi}{2\ttau_c}\frac{\der\psi}{\der\xi}=
    -\f{1}{2\pi}\frac{\der}{\der\xi}\int_{-\infty}^{\infty}
    \frac{|\psi(\xi')-\psi(\xi)|^2
    [\psi(\xi')-\psi(\xi)]}
    {(\xi'-\xi)^2}\,\der{\xi'} \,.
\end{equation}
The claim is that this equation describes the blow up of wave slope on vorticity interfaces, and that almost all initial conditions are attracted to this blow up solution in finite time.

We have not been able to solve \eqref{sseq} directly, due to its nonlinear and non-local character. However, we next provide evidence that the observed behaviour in figure \ref{fig:circle_evo} is consistent with the self-similar form proposed in \eqref{ssform}. To this end, starting from a guess for the constant $\mu$, we define
\begin{equation}
\label{sspsi}
    \psi(\xi)=\frac{\int_{\ttau_1}^{\ttau_2}
    \delta^{-\im\mu-\shalf}\cA(\theta_{\max}+\delta^{\shalf}\xi,\ttau)\der\ttau}
    {\int_{\ttau_1}^{\ttau_2}\delta^{-\shalf}\der\ttau}
\end{equation}
as a guess for the form of $\psi$ (here integrals over time are discretised as sums). We use $\delta^{-\shalf}$ to preferentially weight the data closer to the singularity time. We then seek the constant $\mu$ which minimises
\begin{equation}
\label{ssH}
    H(\mu)=\frac{\int_{\ttau_1}^{\ttau_2}
    \delta^{-\shalf}\int_{-\xi_{\max}}^{\xi_{\max}}w(\xi)|\delta^{-\im\mu}\cA(\theta_{\max}+\delta^{\shalf}\xi,\ttau)-\psi(\xi)|^2\der\xi\der\ttau}
    {\int_{\ttau_1}^{\ttau_2}\delta^{-\shalf}\der\ttau\int_{-\xi_{\max}}^{\xi_{\max}}w(\xi)\der\xi}
\end{equation}
where $w(\xi)=1+\cos(\pi\xi/\xi_{\max})$ is a spatial weight favouring values in the centre of the interval (and vanishing at the endpoints). Again, spatial integration is done by discrete sums (here over 1024 intervals in $\xi$). In practice, $H(\mu)$ exhibits a single, simple minimum at the target value of $\mu$.

\begin{figure}
    \centering
    \includegraphics[width=0.625\textwidth]{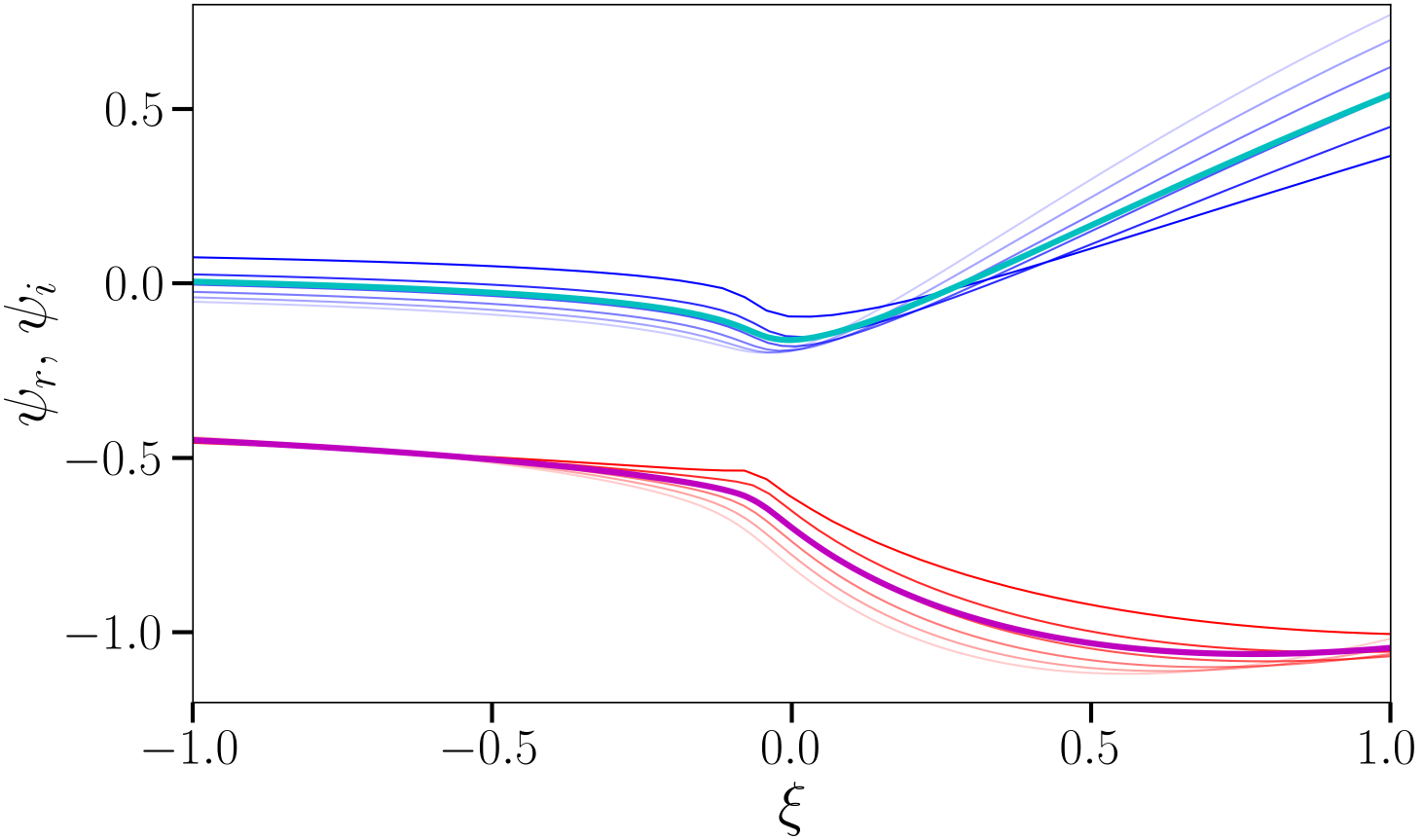}
    \caption{The estimated self-similar solution $\psi(\xi)$ (real part in cyan, imaginary part in magenta), together with scaled numerical profiles (see text) at times $\ttau=0.345$, $0.347$, $0.349$, $0.351$, $0.353$ and $0.355$ (real part in blue, imaginary part in red, with fading backwards in time). [colour online]}
    \label{fig:circle_selfsim}
\end{figure}

We have chosen the time interval between $\ttau_1=0.345$ and $\ttau_2=0.355$ (including both endpoints) to do the analysis, because this interval provides the best fit to the observed wave slope growth in figure \ref{fig:circle_tfit}. We have also chosen $\xi_{\max}=1$ to limit periodicity effects coming from the range in $\theta$ when sampling $\cA$ in \eqref{sspsi} and \eqref{ssH}.  With these choices, we find $\mu\approx0.2307$ where $H(\mu)\approx0.005163$. For this value of $\mu$, we can construct $\psi(\xi)$ from \eqref{sspsi} and compare this with the scaled numerical profiles, specifically with $\delta^{-\im\mu}\cA(\theta_{\max}+\delta^{\shalf}\xi,\ttau)$, at selected times in the sampling period.  This is done in figure \ref{fig:circle_selfsim}. While the collapse is not perfect, it is suggestive of the existence of a self-similar solution, $\psi(\xi)$. The neglect of terms of $\mathcal{O}(\delta^{\shalf})$ likely results in some of the scatter observed. For instance, the neglected term in \eqref{partial} involving $\der\theta_{\max}/\der\ttau$ is questionable, since this derivative is nearly $-8$ just before the singularity time, whereas $\delta^{\shalf}$ varies from $0.1117$ to $0.04983$ over the time period analysed. This neglected term is not necessarily small. Nonetheless, the results in figure \ref{fig:circle_selfsim} at least provide a hint to the possible form of the solution to \eqref{sseq}.

\subsection{The line}

Equation \eqref{aeql} governing the onset of filamentation on the periodic line is nearly identical to that on the sphere, \eqref{aeq2}, except for the curvature term $|\cA|^2\cA$ in the definition of $\cB$, and a sign difference.  In fact, the complex conjugate of $\cA$ for the line satisfies the same equation as $\cA$ for the sphere, after dropping the curvature term.  Hence, to compare the sphere and line most closely, we use the same initial condition (after complex conjugation) in spectral form, namely $a_2(0)=\sqrt{3}/2$ and $a_3(0)=0.25e^{-\im\pi/3}$, with all other coefficients zero.

\begin{figure}
    \centering
    \includegraphics[width=\textwidth]{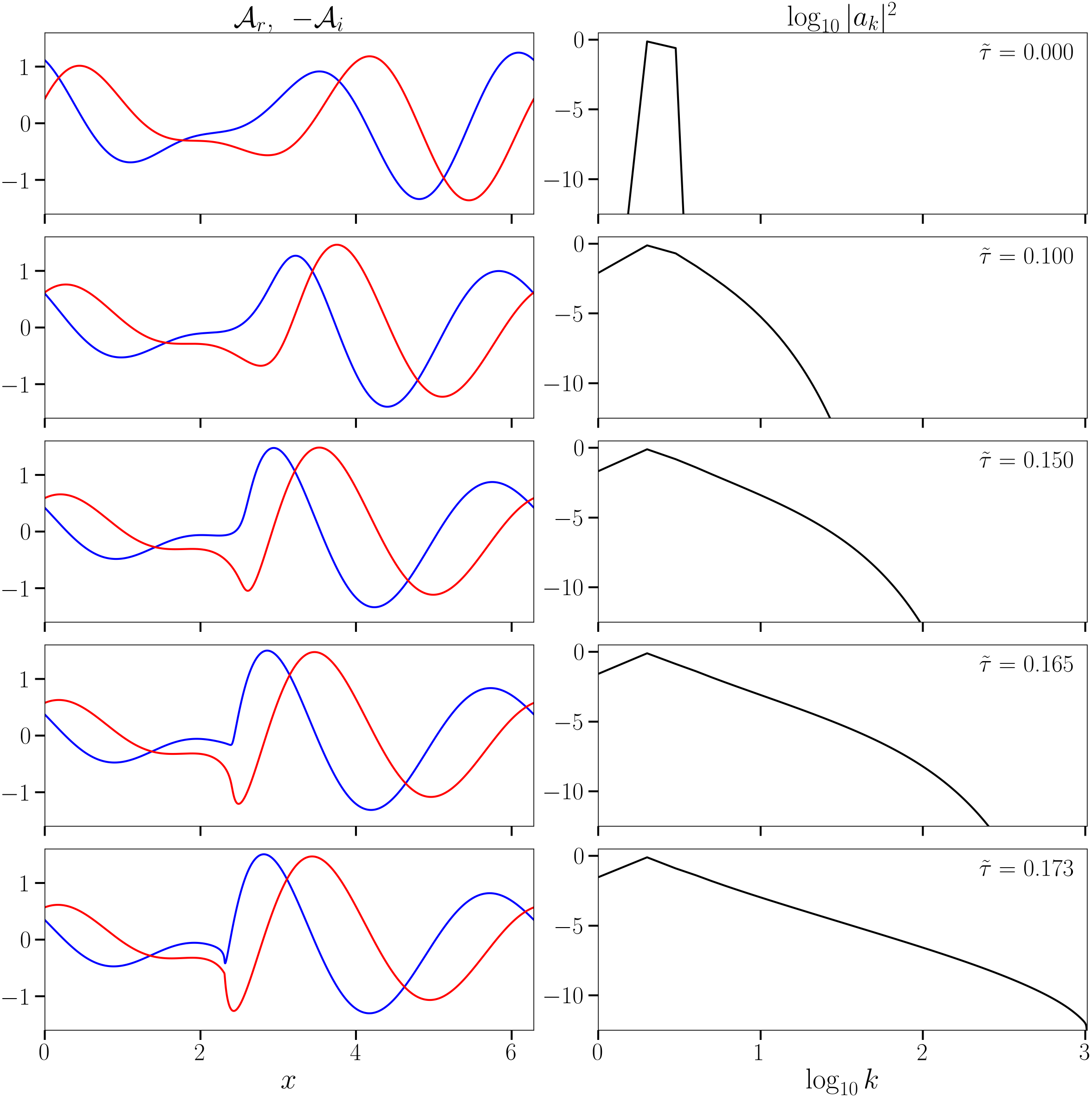}
    \caption{Time evolution of the wave amplitude $\cA$ (left, with real and imaginary parts in blue and red respectively) for a vorticity interface on the periodic line, together with the corresponding power spectrum $|a_k|^2$ (right) for 5 selected times $\ttau$ (increasing downwards).  Initially, just $a_2$ and $a_3$ are non-zero. Compare with figure \ref{fig:circle_evo} for the circular case. [colour online]}
    \label{fig:line_evo}
\end{figure}

\begin{figure}
    \centering
    \includegraphics[width=0.625\textwidth]{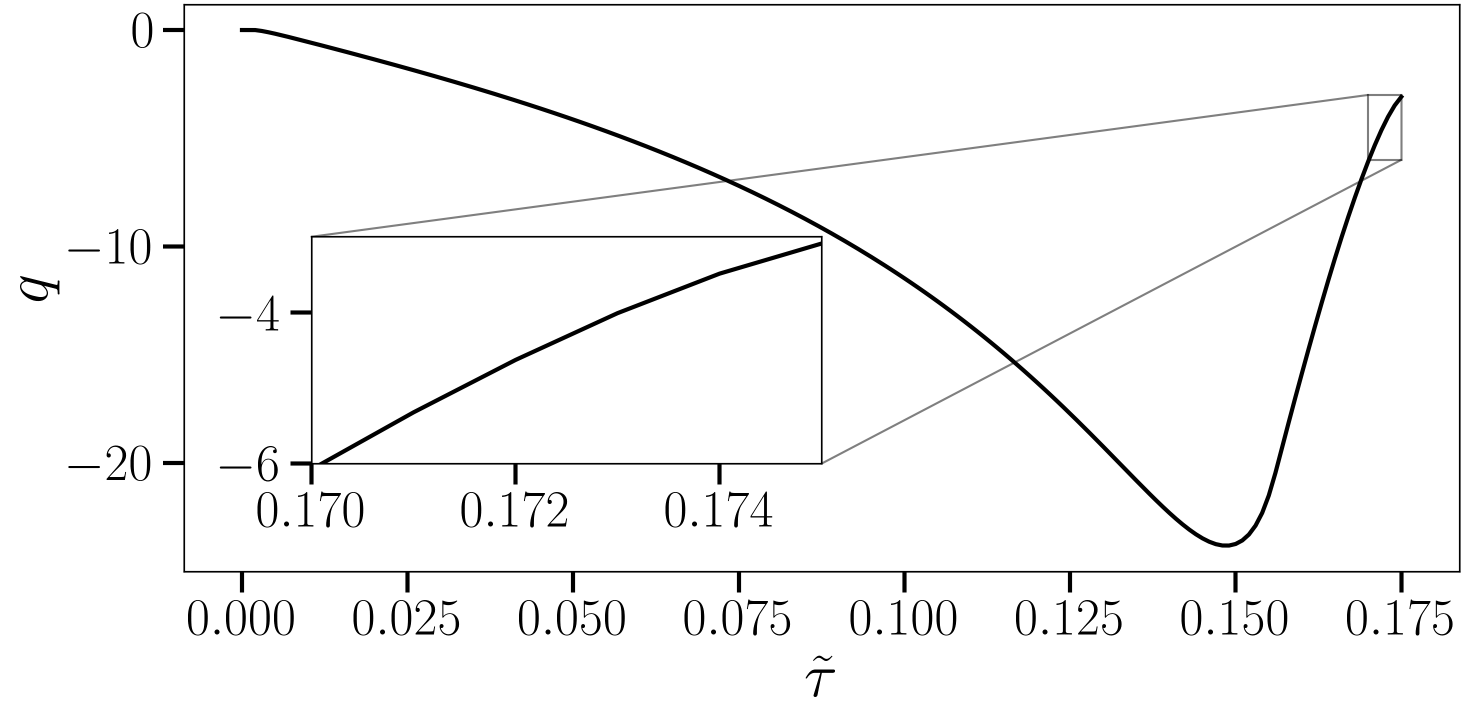}
    \caption{Time evolution of the spectral slope $q$, obtained by a least-squares fit of $\log|a_k|^2$ to $q\log{k}+c$, between wavenumbers $k=10$ and $M/2=1024$.  The inset shows the slope over the last $0.1$ units of time computed (note, the results are not considered reliable beyond $\ttau=0.173$ due to insufficient resolution). Compare with figure \ref{fig:circle_slope} for the circular case.}
    \label{fig:line_slope}
\end{figure}

The evolution of $\cA$ and of its spectrum $|a_k(\ttau)|^2$ are shown at a few selected times in figure \ref{fig:line_evo}. As for the sphere, the waves generally propagate to the left and steepen. The spectrum develops a powerlaw form $\propto k^{q}$, with the exponent $q$ tending to $-3$ at the latest time --- see figure \ref{fig:line_slope}. The same behaviour was found for the sphere. At the final reliable time shown ($\ttau=0.173$), the amplitude $\cA$ exhibits a near discontinuous variation. However, the details differ somewhat from those seen for the sphere in figure \ref{fig:circle_evo}, for instance the `kink' now appears in the real part of $\cA$ rather than the imaginary part. Also, the evolution is nearly twice as fast. Finally, unlike for the sphere, the lowest harmonic $a_1$ varies in time.

\begin{figure}
    \centering
    \includegraphics[width=\textwidth]{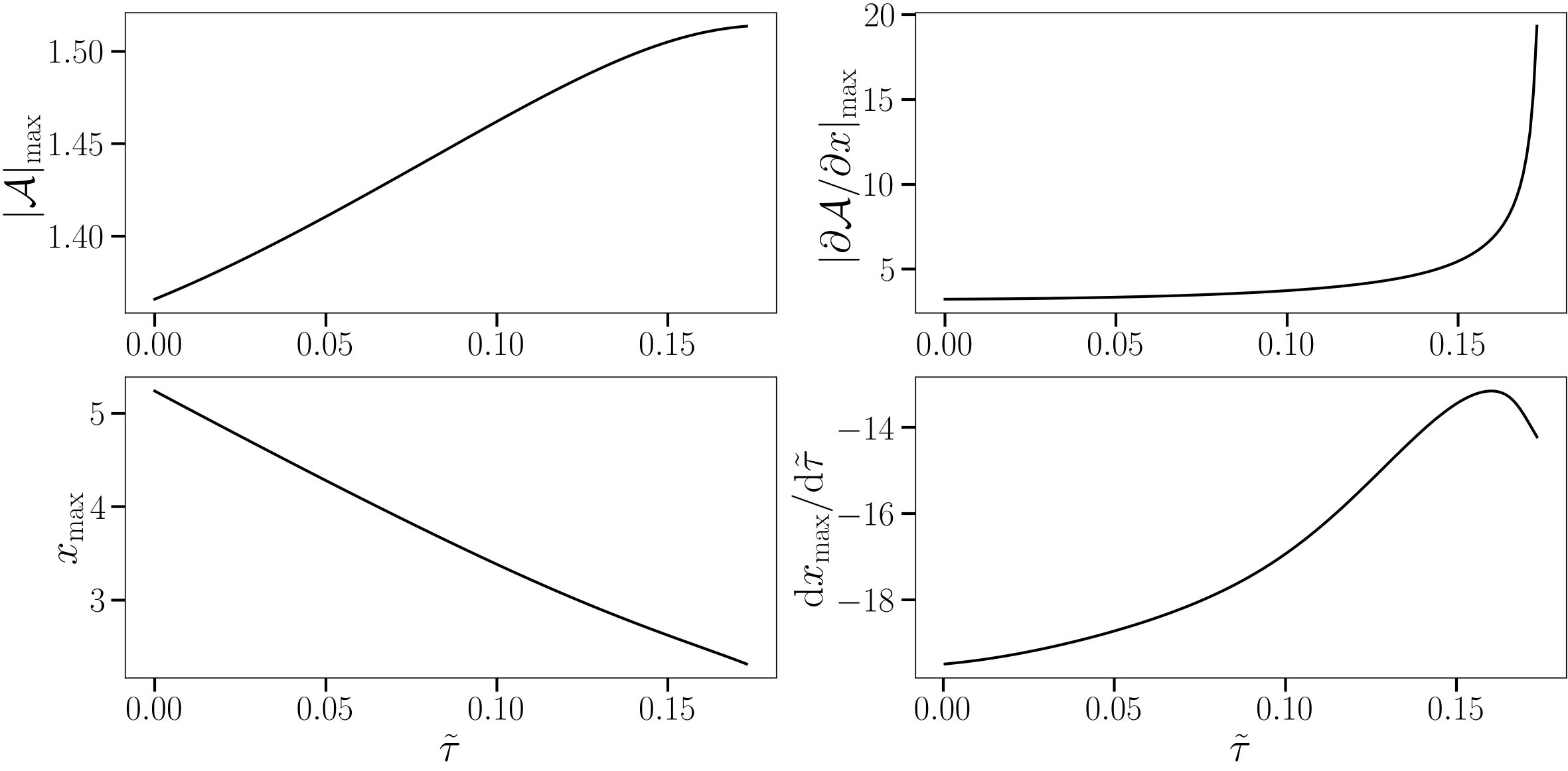}
    \caption{Time evolution of various diagnostics, as labelled, from $\ttau=0$ to $0.173$ (the last reliable time). In the figure for $\der{x}_{\max}/\der{t}$, 1-2-1 averages were repeated 8 times to remove most of the noise occurring around the maximum near $\ttau=0.16$. Compare with figure \ref{fig:circle_diag} for the circular case.}
    \label{fig:line_diag}
\end{figure}

We next examine the evolution of the maximum wave slope $|\partial\cA/\partial{x}|_{\max}$ in figure ~\ref{fig:line_diag}, along with other diagnostics. There is again an indication that the system is approaching a finite-time singularity in wave slope. Near the final time shown, the maximum wave amplitude $|\cA|_{\max}$ hardly varies. The location of maximum wave slope, $x_c(\ttau)$, moves at a moderate speed, even faster than in the case of the sphere (indeed almost twice as fast).

\begin{figure}
    \centering
    \includegraphics[width=\textwidth]{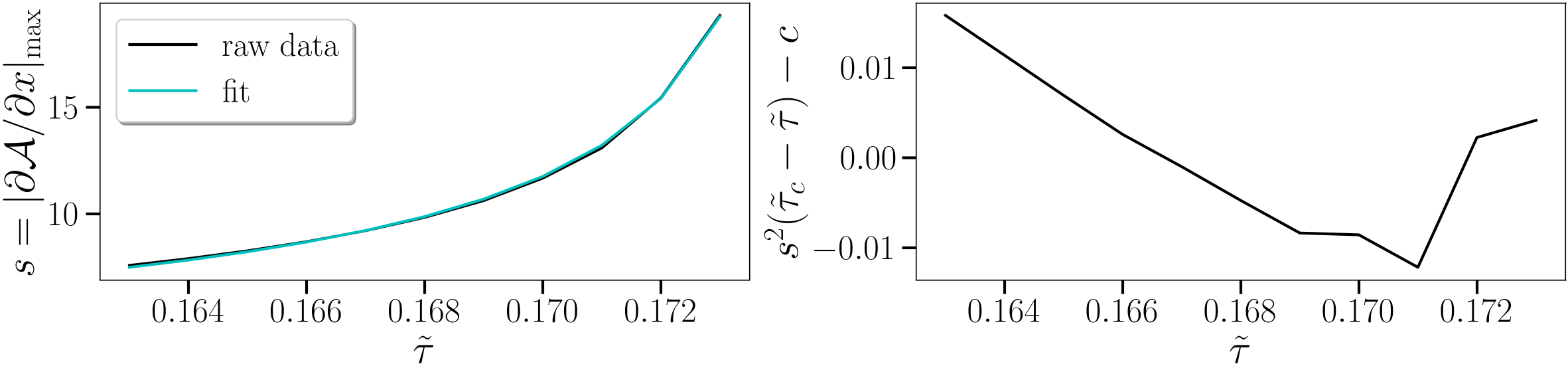}
    \caption{Time evolution of the maximum wave slope $s(\ttau)=|\partial\cA/\partial{x}|_{\max}$ together with a fit to $\sqrt{c/(\ttau_c-\ttau)}$ (left), and the function $f(\ttau)=s^2(\ttau_c-\ttau)-c$ (right) which would be zero for a perfect fit. Compare with figure \ref{fig:circle_tfit} for the circular case. [colour online]}
    \label{fig:line_tfit}
\end{figure}

Like in the case of the sphere, the maximum wave slope $s(\ttau)=|\partial\cA/\partial{x}|_{\max}$ appears to exhibit a square-root singularity $s\sim\sqrt{c/(\ttau_c-\ttau)}$, now for $\ttau_c\approx0.17478$, as shown in figure \ref{fig:line_tfit}. The singularity time $\ttau_c$ and the constant $c$ were determined by a least-squares fit over the period $0.163\leq\ttau\leq0.173$, using the same approach used for the sphere. The r.m.s.\ error in the fit over this period is only $0.00788$ (note: $c\approx0.65997$). 

\begin{figure}
    \centering
    \includegraphics[width=0.625\textwidth]{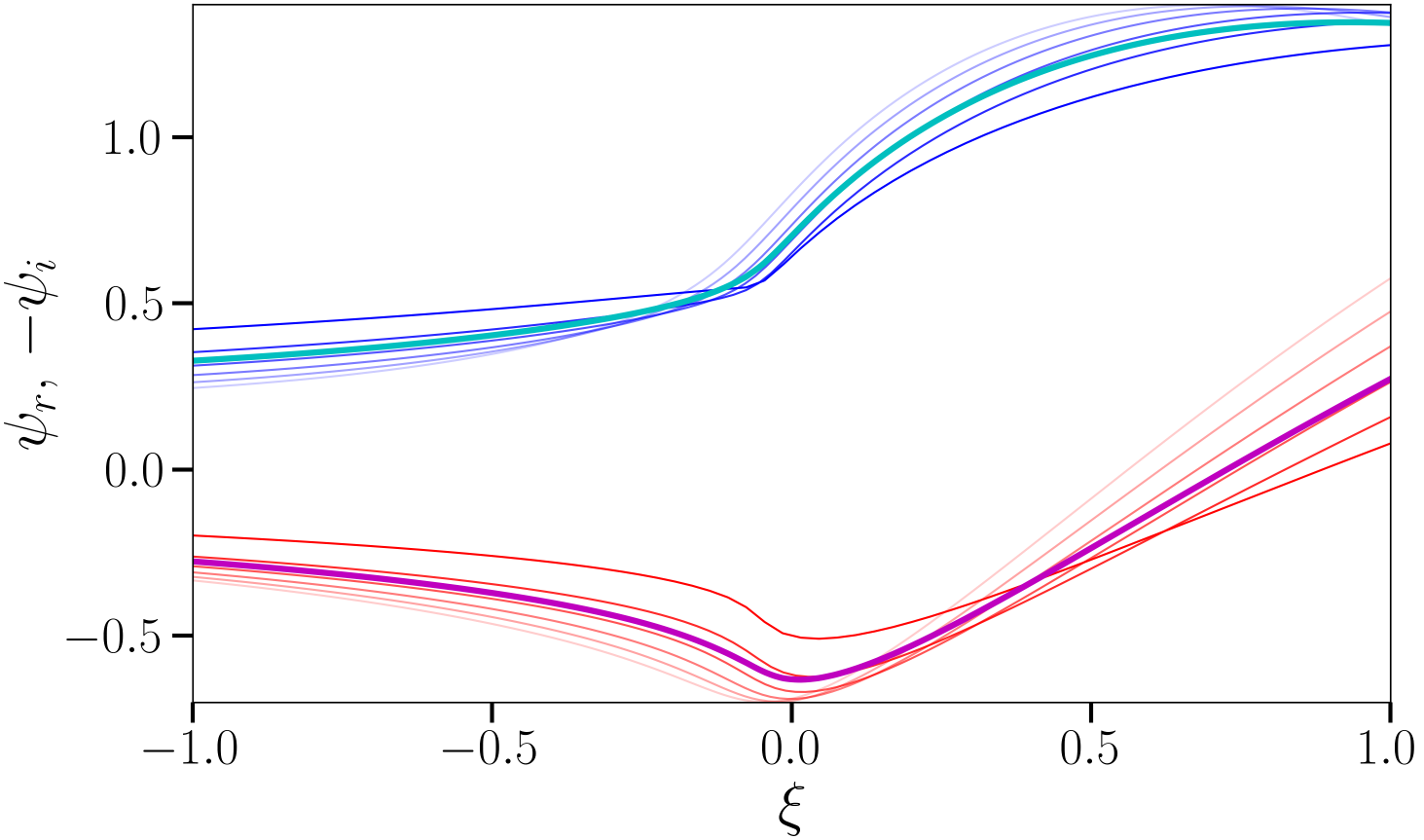}
    \caption{The estimated self-similar solution $\psi(\xi)$ (real part in cyan, imaginary part in magenta), together with scaled numerical profiles (see text) at times $\ttau=0.163$, $0.165$, $0.167$, $0.169$, $0.171$ and $0.173$ (real part in blue, imaginary part in red, with fading backwards in time). Compare with figure \ref{fig:circle_selfsim} for the circular case. [colour online]}
    \label{fig:line_selfsim}
\end{figure}

We next provide evidence for a self-similar solution of the form $\cA=\delta^{\im \mu} \psi(\xi)$, where $\xi$ and $\delta$ have the same definitions given in \eqref{ssvar} except $\theta$ is replaced by $x$. One can derive an equation for $\psi$ analogous to that given in \eqref{sseq}, and indeed the only difference is that the complex conjugate of $\psi$ satisfies that equation for the line. As for the sphere, we fit the scaled numerical data over a time period extending from $\ttau_1=0.163$ to $\ttau_1=0.173$ to estimate the form of $\psi(\xi)$ and the exponent $\mu$. That fit gives $\mu\approx0.3150$ (cf.\ $\mu\approx0.2307$ for the sphere), and  $H(\mu)\approx0.005163$ in \eqref{ssH}. The estimated form of $\psi$ along with the scaled numerical data $\delta^{-\im\mu}\cA(x_{\max}+\delta^{\shalf}\xi,\ttau)$ are shown in figure \ref{fig:line_selfsim}. The form of $\psi$ compares well with the spherical case in figure \ref{fig:circle_selfsim}, though there are some differences. These differences, and the scatter in the fit, may result from the impact of higher-order terms, formally $\mathcal{O}(\delta^{\shalf})$ smaller but in practice potentially comparable due to the computational difficulty in getting closer to the singularity time. Nonetheless, these results hint at the likely form of the solution to the self-similar equation \eqref{sseq}.

\subsection{Stability of travelling waves}

\begin{figure}
    \centering
    \includegraphics[width=\textwidth]{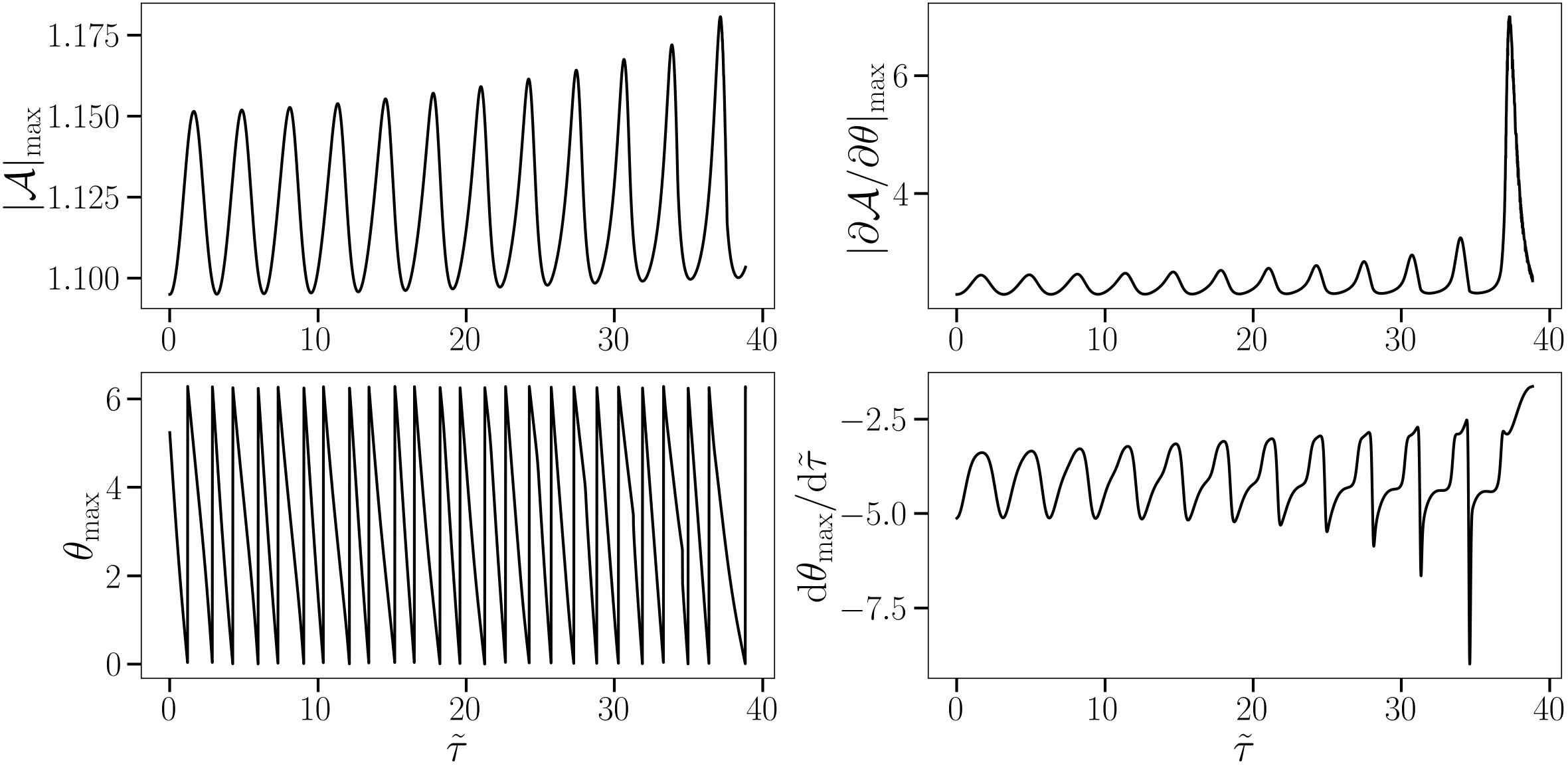}
    \caption{Time evolution of various diagnostics for a simulation starting from a 
    weakly-perturbed travelling wave. See figure \ref{fig:circle_diag} for the analogous diagnostics in the case of a larger initial perturbation.}
    \label{fig:m2diag}
\end{figure}

All single-mode disturbances, which take the form
\[
\cA(\theta,\ttau)=a_k e^{\im k\theta-k(k-1)|a_k|^2\ttau}\,,
\]
are exact solutions of \eqref{aeq2} or \eqref{eqaspec}, and translate in $\theta$ at speed $-(k-1)|a_k|^2$ \citep{Constantin2024}. (In the case of the line, the $k-1$ factor is replaced by $k$.)  Here, we briefly examine the stability of a few of these solutions numerically. We initialise as before by choosing another mode $p>k$ with initial amplitude $a_p(0)=\epsilon e^{\im \pi/3}$, then chose $a_k(0)=\sqrt{1-\epsilon^2}$ so that $\cP=\sum_m |a_m|^2=1$, without loss of generality.  Again, we take $k=2$ and $p=3$, but take $\epsilon=0.1$, which is 5 times smaller than in the previous example. (Other values are discussed below.)

\begin{figure}
    \centering
    \includegraphics[width=0.92\textwidth]{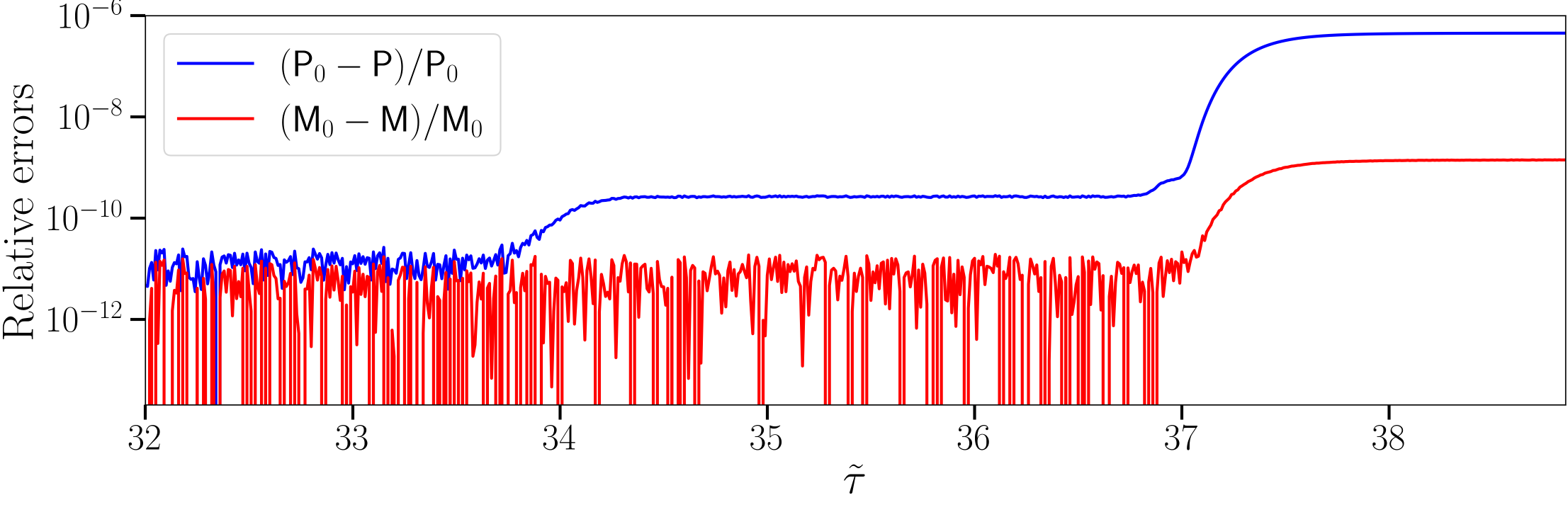}
    \caption{Late time evolution of the relative errors in momentum $\cP$ and mass $\cM$. A subscript `0' refers to their initial values. Prior to $\ttau\approx34$, errors are $\mathcal{O}(10^{-11})$ but likely smaller because the spectral coefficients $a_m$ were only saved to 11-digit accuracy. [colour online]}
    \label{fig:m2errors}
\end{figure}

Figure \ref{fig:m2diag} shows various diagnostics including the maximum wave slope $|\partial\cA/\partial\theta|_{\max}$. Comparing to the case with $\epsilon=0.5$ in figure \ref{fig:circle_diag}, we see that smaller $\epsilon$ delays the onset of filamentation, which appears to occur after a series of growing oscillations (this behaviour is found also for $\epsilon=0.11$ and $0.12$, whereas $\epsilon=0.13$ and larger $\epsilon$ exhibit a single diverging peak in wave slope). The oscillations correspond to the almost periodic amplification and reduction of the tail of the power spectrum, with each amplification being larger than the last.  The wave form $\cA$ begins to develop a kink and a near discontinuity by $\ttau=37.2$. The recovery of $|\partial\cA/\partial\theta|_{\max}$ after this time is likely to be spurious --- significant errors in the conserved momentum $\cP$ and mass $\cM$ develop especially around $\ttau=37$, when the final peak occurs --- see figure \ref{fig:m2errors}. The numerical filtering at high wavenumbers $m$ arrests the growth in wave slope, which likely diverges soon after $\ttau=37$ in reality. Notably, the best-fit spectral slope $q$ climbs through $-3$ at $\ttau=37.06$, when strong numerical dissipation occurs. All evidence points to a wave slope singularity in finite time; it appears that even weakly perturbed translating waves are eventually subject to filamentation.

\section{Discussion}
\label{sec:discussion}

This paper has revisited a generic aspect of vorticity interfaces, namely the tendency for unsteady disturbances to steepen and `break', resulting in `filamentation' \citep{Dritschel1988f}. Specifically, we have studied an equation first derived in \cite{Dritschel1988f} describing the weakly-nonlinear development of shallow (small wave slope) disturbances to circular and linear interfaces. That equation is shown to have a simpler, universal form in a rescaled slow time variable \citep[see also][for details of its mathematical structure, invariants and some exact solutions]{Constantin2024}.

We have studied the onset of filamentation on both a circular interface (applicable to interfaces on the sphere or on the plane) and a periodic linear interface.  In both cases, we find generically that unsteady initial conditions tend to steepen, apparently exhibiting an inverse square root time singularity in wave slope at one location.  This is the manifestation of filamentation discussed in \cite{Dritschel1988f}, who showed that the results compare well with full contour dynamics simulations up to the point of filamentation (this is confirmed here with more accurate simulations). 

Motivated by the numerical findings, we have developed a theory for the self-similar evolution of the interface close to the singularity in wave slope. This is a cubically-nonlinear, nonlocal equation for a complex function $\psi(\xi)$ in a similarity variable $\xi$, of the form that appears in the heat equation. The equation has an eigenvalue $\mu$ related to the time-scaling factor with exponent ${\im \mu}$ used to define $\psi$. We have not been able to solve this equation, but have provided an estimate of the form of $\psi$ by fitting numerical data close to the singularity time. Future work will target finding numerical solutions of this self-similar equation and, more generally, rigourously proving the self-similar equation is an attractor for almost all initial conditions. 

\backsection[Acknowledgements]{
DGD gratefully acknowledges support from the University of Vienna for two research visits.}

\backsection[Funding]{}

\backsection[Declaration of interests]{The authors report no conflict of interest.}


\backsection[Author ORCID]{
D.\ G.\ Dritschel, https://orcid.org/0000-0001-6489-3395;
P.\ M.\ Germain, https://orcid.org/0000-0003-3148-4127;
A.\ Constantin, https://orcid.org/0000-0001-8868-9305.
}

\backsection[Author contributions]{D.\ G.\ Dritschel: conceptualisation, methodology, literature research, software, numerical analysis, writing; P.\ M.\ Germain: conceptualisation, methodology, writing; A.\ Constantin: conceptualisation, methodology, writing}

\bibliographystyle{jfm}
\bibliography{filam.bib}

\end{document}